\documentclass[aps,prl,twocolumn]{revtex4}

\usepackage{graphicx}
\usepackage{color}
\usepackage{dcolumn}
\usepackage{bm}

\begin{document}

\title{Non-equilibrium AC Stark effect and long-lived exciton-polariton states in
  semiconductor Mie resonators}

\author{Andreas Lubatsch$^1$ and Regine Frank$^{2,3,4}$}

\email[Correspondence should be addressed to:]{r.frank@uni-tuebingen.de}

\affiliation{$^1$ Georg-Simon-Ohm University of Applied Sciences, Ke{\ss}lerplatz 12, 90489 N\"urnberg, Germany\\
$^2$ Institute of Theoretical Physics, Optics and Photonics, Eberhard-Karls
University, Auf der Morgenstelle 14, 72076 T\"ubingen, Germany\\
$^3$ Bell Laboratories, 600 Mountain Avenue, Murray Hill, NJ 07974-0636 USA\\
$^4$ Institute of Solid State Physics (TFP), Karlsruhe Institute of Technology (KIT), Wolfgang-Gaede Stra{\ss}e 1, 76131 Karlsruhe, Germany}

\date{\today}

\begin{abstract}
We present Floquet-Keldysh non-equilibrium dynamical mean field calculations
(DMFT) with an iterative perturbative solver (IPT) for high pump-power induced
Stark states yielding a closing gap
within semiconductor bulk and nano-cavities.  Our model predicts
unusual broadening of bands in systems like ZnO random lasers. This can be
explained as an exciton-polariton
coupling within the single crystalline pillar meeting the Mie resonance
condition. Extraordinary enhancement of the lifetime of electronic states
within the gap can lead to stable lasing modes and gain narrowing.
\end{abstract}

\maketitle

The research of random and polariton lasing has fascinated the community ever
since their discovery \cite{Cao,YaNature,YaScience,Imamoglu,Dang,
  Hoefling,Li,Guillet,Bajoni,Wouters}. 
Both phenomena of coherent emission originating from
semiconductor micro-resonators have been interpreted as to be fundamentally
different. We show in this letter that on the basis of ultrafast
non-equilibrium excitation the groundbreaking implication is possible to
interpret random lasing as coupled cavity polariton lasers. On the other hand
it would also allow to deduce that semiconductor cavity polariton lasers
develop laser thresholds which is the signature of a phase transition.

\begin{figure}[t]
\vspace*{0.5cm}\rotatebox{0}{\scalebox{0.35}{\includegraphics{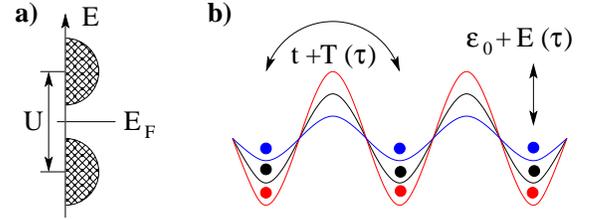}}}
\vspace{0.0cm}
\caption{Insulator to metal transition. (a) Split energy band. The local Coulomb interaction $U$
  determines the width of the gap symmetrical to the Fermi energy $E_F$. (b) Electrons in the crystal lattice
  structure (black). Periodic driving yields an
  additional hopping contribution $T(\tau)$ and local renormalization of the
  lattice energy $\epsilon_k = \epsilon_0 +E(\tau)$. The colors represent three time
  steps of driving.}
\label{Mott}
\end{figure}
\begin{figure}[t]
\vspace*{-0.0cm}\hspace*{-0.0cm}\rotatebox{0}{\scalebox{0.19}{\includegraphics[clip]{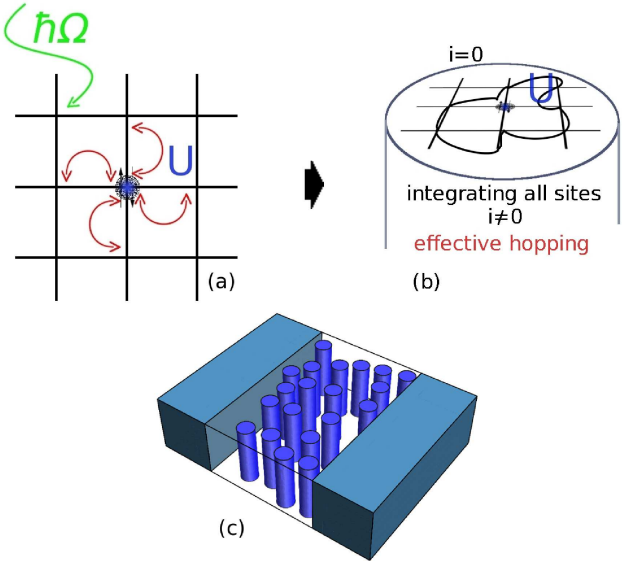}}}
\vspace{-0.0cm}
\caption{(a) The
  semiconductor behaves in this regime as an insulator: In the
  non-equilibrium DMFT-scheme optical excitation yields
  electronic hopping processes. They are mapped to the interaction with the single site on the background of
  the surrounding lattice (bath). (b) Integration over all sites leads to
  an effective theory including non-equilibrium excitation. This approach here
  is self-consistent because the bath consists of these very same single sites. 
  The driven electronic system is additionally
   coupled to a cavity-resonance, the Mie
  resonance. (c) Experimentally relevant setup, a disordered waveguide of ZnO
  nanostructures embedded in a Si waver.}
\label{DMFT}
\end{figure}

In this letter we consider the semiconductor bulk as insulator, which is
true in the zero temperature limit ($T=0$), where all lattice influences are
frozen out. The gap of $3.3\,eV$ of ZnO corresponds to a Coulomb repulsion $U$
of about $U/D=4.0$, which is deeply in the Mott insulating regime. D is the half bandwidth.
First, the non-equilibrium bandstructure (i) of a classically strong field
pumped bulk ZnO semiconductor is investigated by Keldysh-Floquet
non-equilibrium dynamical mean field theory (DMFT) and solved using second
order non-equilibrium iterative perturbation theory (IPT). 
Second we explore topological effects (ii) induced in the non-equilibrium bandstructure 
when it is coupled to the
cavity mode of a geometrical resonator (Fig. \ref{DMFT}). 
By comparison of the life-times of both
non-equilibrium solutions we identify long-lived resonances, exciton-polaritons, in the cavity coupled
solutions which match in the frequency range experimentally determined spectra
of both polariton lasers in ZnO nano-pillars as well as ZnO random lasing
experiments.\\
The Hubbard-type Hamiltonian
  Eq. (\ref{Hamilton_we}) describes the electronic dynamics responsible for
  the generation of microscopic gain and thus leading in further
  consequence to electronic inversion. It is generalized for
electron-photon interaction
which is also solved by applying the Floquet-Keldysh formalism using dynamical mean
field theory (DMFT) and an iterative perturbative solver (IPT) (see Fig. \ref{Mott}). 
The classically field driven cavity-coupled Hamiltonian in position space
reads

\begin{eqnarray}
\!\!\!\!\!\!\!\!H\!&=&\!\! \sum_{k, \sigma} \! \epsilon_k  c^{\dagger}_{k,\sigma}c^{{\color{white}\dagger}}_{k,\sigma} 
+   \frac{U}{2} \sum_{i, \sigma} c^{\dagger}_{i,\sigma}c_{i,\sigma}c^{\dagger}_{i,-\sigma}c_{i,-\sigma}\label{Hamilton_we}
\\&& - t\!\! \sum_{\langle ij \rangle, \sigma}\!\!
c^{\dagger}_{i,\sigma}c^{{\color{white}\dagger}}_{j,\sigma}\nonumber
\\&& + i\vec{d}\cdot\vec{E}_0 \cos(\Omega_L \tau)\sum_{<ij>} 
 \left(
           c^{\dagger}_{i,\sigma}c^{{\color{white}\dagger}}_{j,\sigma} 
 	  -
           c^{\dagger}_{j,\sigma}c^{{\color{white}\dagger}}_{i,\sigma} \right)\nonumber
\!\\&& + \hbar\omega_o  a^{\dagger}a^{{\color{white}\dagger}} + g\! \sum_{k, \sigma}
c^{\dagger}_{k,\sigma}c^{{\color{white}\dagger}}_{k,\sigma}  (a^{\dagger}  \! +
a)  .\nonumber
\end{eqnarray}
The first term $\sum_{k, \sigma} \! \epsilon_k
c^{\dagger}_{k,\sigma}c^{{\color{white}\dagger}}_{k,\sigma}$ denotes the onsite processes which is 
completed by $\frac{U}{2} \sum_{i, \sigma} c^{\dagger}_{i,\sigma}c_{i,\sigma}c^{\dagger}_{i,-\sigma}c_{i,-\sigma}$ devoted
to the Coulomb interaction $U$ between electrons with opposite spins. 
The third
term $-t \sum_{\langle ij \rangle, \sigma}\!\!
c^{\dagger}_{i,\sigma}c^{{\color{white}\dagger}}_{j,\sigma}$ denotes the
hopping processes with the amplitude $t$ between nearest neighbored sites. 
The last three terms are
devoted to photons, including the external field-electron coupling, 
the kinetic energy of the cavity photons and the cavity photon-electron interaction. 
In $ i\vec{d}\cdot\vec{E}_0 \cos(\Omega_L \tau)\sum_{<ij>} 
 \left(
           c^{\dagger}_{i,\sigma}c^{{\color{white}\dagger}}_{j,\sigma} 
 	  -
           c^{\dagger}_{j,\sigma}c^{{\color{white}\dagger}}_{i,\sigma}
         \right)$ is noted the renormalization of the hopping processes due to classical interaction of the dipole moment $\vec d$ with an external time
         dependent electric
         field of the amplitude $\vec E_0$, the pumping laser. $\tau$ denotes
         its time dependency and $\Omega_L$ the corresponding laser frequency. The last two
         terms denote the cavity mode $\hbar\omega_o
         a^{\dagger}a^{{\color{white}\dagger}}$ at the resonance frequency $\omega_0$ as well as
         the coupling of this resonance to the electronic density of states $g\! \sum_{k, \sigma}
c^{\dagger}_{k,\sigma}c^{{\color{white}\dagger}}_{k,\sigma}  (a^{\dagger}  \! +
a)$
         with coupling strength $g$. The creator
         (annihilator) of a cavity photon is denoted $a\,(a^\dagger)$ and of an
         electron it is $c\,(c^\dagger)$. The subscript $k$ denotes the site, $i,j$ denote next neighbored sites (Fig. \ref{Mott}) within the
         originally assumed single-band. The first four terms of Hamiltonian Eq. (\ref{Hamilton_we})
describe (i) externally pumped bulk semiconductor.\\
The explicit time dependence of the external field yields Green's functions that depend
on two separate time arguments. A double Fourier
transform from time- to frequency space leads to relative and
center-of-mass frequency \cite{PRB,FrankANN}
\begin{eqnarray}
\label{Floquet-Fourier}
G_{mn}^{\alpha\beta} (\omega)
\!\!\! &=&\!\!\!
\left\lmoustache \!\!{\rm d}{\tau_1^\alpha}\!{\rm d}{\tau_2^\beta}\right.
e^{-i\Omega_L(m{\tau_1^\alpha}-n{\tau_2^\beta})}
e^{i\omega({\tau_1^\alpha}-{\tau_2^\beta})}
G (\tau_1^\alpha,\tau_2^\beta)\nonumber\\
\!\!\!&\equiv&\!\!\!
G^{\alpha\beta} (\omega-m\Omega_L, \omega - n\Omega_L).
\end{eqnarray}
$(m,n)$ label the Floquet modes and $(\alpha, \beta)$ address the branch
of the Keldysh contour ($\pm$) where the respective time argument resides. 
Floquet modes describe the Fourier-transformation of a
periodic potential in time as the principal structure of bands in frequency space. However the
physical consequence which is noteworthy is a quantized absorption and
emission of energy quanta
$\hbar \Omega_L$ of and to the original classical field. For non-interacting electrons the analytical solution for 
$G_{mn}(k,\omega) $ is solving the equation of motion. Photo-induced hopping
leads to the retarded Green's function for this sub-system 
\begin{eqnarray}
G_{mn}^{R}(k,\omega) \label{Green}
=
\sum_{\rho}
\frac
{
J_{\rho-m}\left(A_0\tilde{\epsilon}_k \right)
J_{\rho-n}\left(A_0\tilde{\epsilon}_k \right)
}
{
\omega -\rho\Omega_L - \epsilon_k + i 0^+
},
\end{eqnarray}
where $\tilde{\epsilon}_k$ represents the dispersion relation induced by the
external field and is to be distinguished from $\epsilon$ the lattice dispersion.
The $J_n$ are the cylindrical Bessel functions of integer order, 
 $A_0 = \vec{d}\cdot\vec{E}_0 $ and
$\Omega_L$ is the laser frequency.
The physical Green's function is given according to
\begin{eqnarray}
\label{EqGsum}
G^{R}_{\rm Lb}(k,\omega)  
=
\sum_{m,n}
G_{mn}^{R}(k,\omega).
\end{eqnarray}
The Hamiltonian is solved through single-site Floquet-Keldysh DMFT (IPT) \cite{NJP,ANN}. This
requires strict periodicity of the lattice and homogeneous driving. That
condition implies Bloch states and is characterized by the wave vector k within
the laser-band-electron Green's function $G^{R}_{\rm Lb}(k,\omega)$. Due to
these requirements and the additional cavity in our system the DMFT selfconsistency relation, the centerpiece of DMFT, assumes the form of a matrix equation
of dimension $2 \times 2$ in Keldysh space of the nonequilibrium Green's
functions and of dimension $n \times n$ in Floquet space. As DMFT solver we
generalized the iterated perturbation theory (IPT) to Keldysh-Floquet form as well. The resulting numerics, although 
challenging, proves to be efficient and stable also for all values of the
Coulomb interaction $U$. Whereas we consider in \cite{NJP} a renormalized
lattice potential, we take into account here a coupling of the microscopic
dipole moment to the external field amplitude \cite{PRB,FrankANN}. This implies the
quantum-mechanical character of the dipole operator $\hat d\,=\, i\vec d\, \sum_{<ij>} 
 \left(
           c^{\dagger}_{i,\sigma}c^{{\color{white}\dagger}}_{j,\sigma} 
 	  -
           c^{\dagger}_{j,\sigma}c^{{\color{white}\dagger}}_{i,\sigma} \right)$ in the fourth term of
Eq.(\ref{Hamilton_we}). It is fundamentally different from the generic kinetic
hopping of the third term. The coupling $\hat d \cdot \vec E_0
\cos(\Omega_L\tau)$ under the assumption of the Coulomb gauge $\vec E(\tau)\,=
- \frac{\partial}{\partial \tau} \vec A(\tau)$, which reads in Fourier space
$\vec E(\Omega_L)\,=\,i\Omega_L\cdot\,\vec A(\Omega_L)$, generates the factor
$\Omega_L$ that cancels the $1/\Omega_L$ term in the renormalized cylindrical Bessel function in Eq. (7) of
ref. \cite{NJP}. Consequentially we consider the first $n=10$ Floquet modes only which has
been proven to be sufficient here.

\begin{figure}[t]
\vspace*{0.5cm}\rotatebox{0}{\scalebox{0.35}{\includegraphics{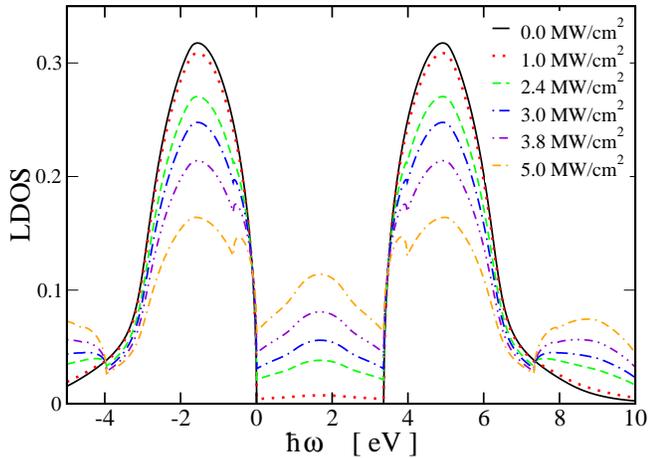}}}
\vspace{0.0cm}
\caption{LDOS for ZnO bulk
  semiconductor. In equilibrium the electronic gap is $3.38 eV$, the microscopic dipole moment equals $|d|=2.30826288 \times 10^{-28} Cm$, the lattice
    constant $a=0,325\, nm$. The pump
  power is increased from $0$ to $5.0 MW/cm^2$. A significant shift of states
  into the semiconductor gap is observed. This energy region has been
  experimentally identified to be the gain area for polaritonic
  states. Corresponding life times can be found in Fig. \ref{IMAG}.}
\label{LDOS}
\end{figure}

The physical consequences are fascinating since in the non-equilibrium
externally pumped state the conventional bulk
semiconductor ZnO exhibits a counter-intuitive behavior. The parameters are chosen such that the resulting band split equals the ZnO
semiconductor band gap of $3.3\,eV$ in the non-pumped case. Through
classical external excitation a multi-band insulator develops, where the
quantized exchange manifests itself in a multitude of sub-bands. The original
bandgap of $3.3\,eV$ is subsequently occupied with dressed states or excitons \cite{SchmittRink}. The complex permittivity is defined as 
$\hat\epsilon = \epsilon\left[1 - i \frac{\sigma}{\omega \epsilon}\right] = \epsilon' - i\epsilon''$ 
where $\epsilon'= \epsilon_r\epsilon_0$ is the real part of the
permittivity. The optical conductivity determines the imaginary part 
$\omega\epsilon'' = \sigma/\omega\epsilon$ 
\begin{eqnarray}
\label{sigma_dc}
\sigma&&\!\!\!\!\!\!(\Omega_L,i\omega)
=
\sum_{m} 
\frac{8e^2t^2}{2\pi^3}\!\!
\int\!\! {\rm d}{\epsilon}
N_0(\epsilon)\\
&&\times
\!\!\int\!\! {\rm d}{\omega \,'}\!
\left({\rm Im\,}G^{R}_{0m}
(\epsilon, \omega \,',\Omega_L)\right)^2 \!\!
\frac{\partial}{\partial \omega\,'}  F^{neq}_{0m}(\omega\,', \Omega_L),
\nonumber
\end{eqnarray}
and characterizes the microscopic gain spectrum (see refs.\cite{Kotliar,NJP}).\\
The optical cavity and the Fano-coupling \cite{Fano} to the electronic
subsystem of the bulk semiconductor is described (ii) in the last term of the original Hamiltonian, Eq.(\ref{Hamilton_we}).
It yields an additional contribution to the self-energy $\Sigma$ in the
denominator of the retarded Green's function Eq.(\ref{Green}). The form of
$\Sigma$ significantly triggers the numerical efficiency of the
selfconsistency consideration. The strength $g$ of the coupling between optically excited electrons $c^{\dagger}_{k,\sigma}c^{{\color{white}\dagger}}_{k,\sigma}$ and the cavity resonance $\hbar\omega_0a^{\dagger}a$ has to be
considered $g<1$ in order to guarantee convergence of the perturbation
series up to second or higher order.
\begin{figure}[t]
\vspace*{0.5cm}\rotatebox{0}{\scalebox{0.35}{\includegraphics{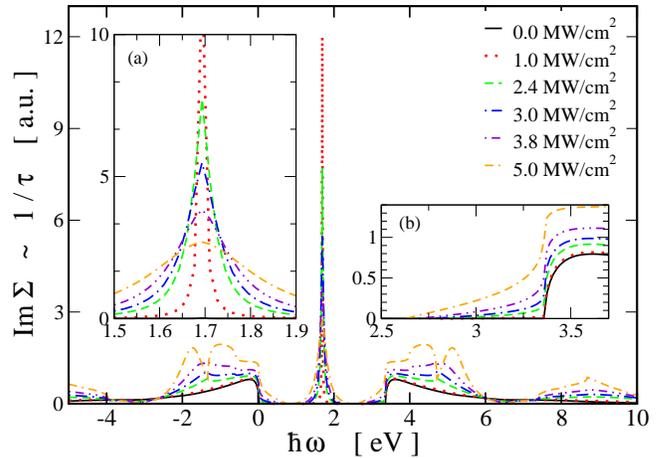}}}
\vspace*{0.0cm}
\caption{Inverse lifetime $1/\tau$ corresponding to the LDOS (Fig. \ref{LDOS}) for increasing pump power. (a)
  The peak centered at $1.69 eV$ indicates small lifetimes, fast decay
  of mid-gap states (red emission). (b) Finite lifetime of non-equilibrium states
  near the band edge indicate stable states which can couple to geometric
  (Mie) resonances and form cavity-polaritons. With rising pump power the
  lifetime drops but remains larger than the lifetime of nearby
  excitonic states in the band.}
\label{IMAG}
\end{figure}
Our results for the non-equilibrium bandstructure (i) of the classically
excited ZnO bulk (Fig. \ref{LDOS}) show for increasing pump strengths
(colored graphs) spectral weight inside the semiconductor band gap whereas
for zero excitation  (black curve) the original semi-circular
bandstructure for non-interacting systems (band-width assumed to be $1.7\,eV$)
splits through the Coulomb
interaction $U$ in upper
and lower band. The tails at the outer upper and lower bandedges are typical
for results derived by the DMFT (IPT) solver. It is found that the additional Floquet bands emerge with increasing
pump intensity. Further the development of moderate sub gaps towards the near-gap
band edges at $-0.67\,eV$ and $4.05\,eV$ (compared to the bandedge of the
valence band) is seen whereas the band edges themselves show a sharp step towards the
novel spectral weight inside the gap. Interestingly the relation of the
sub-gaps and the simultaneously found features in the results for the non-equilibrium lifetimes $1/{\rm Im} \Sigma$ of the excited
states (see Fig. \ref{IMAG}) is non-trivial. We find for the same excitation wavelength
($\lambda\,=\, 710nm$) and a raising intensity the increase of the
non-equilibrium lifetime at the Fermi edge (see
Fig. \ref{IMAG}a). Also $1/\tau$ is larger in the gap near the
band edges which corresponds with a higher local density of electronic states
(LDOS), i.e. a rising
mobility. A plateauing of lifetimes is observed near the
edge, whereas deep in the band at $4.8eV$ we derive for increasing intensity a
crossover. Here $1/\tau$ develops a peak towards $|E_0^2|\,=\,3.8MW/cm^2$
which flips into a sharp dipped feature (red dash-dots) and arrives at a
lifetime almost as high as in the non-excited case (black line). This behavior
is rather unexpected however it is confirmed by a consistency check of the
Floquet sum.

\begin{figure}[t!]
\vspace*{0.0cm}\hspace*{-0.0cm}{\rotatebox{0}{\scalebox{0.35}{\includegraphics[clip]{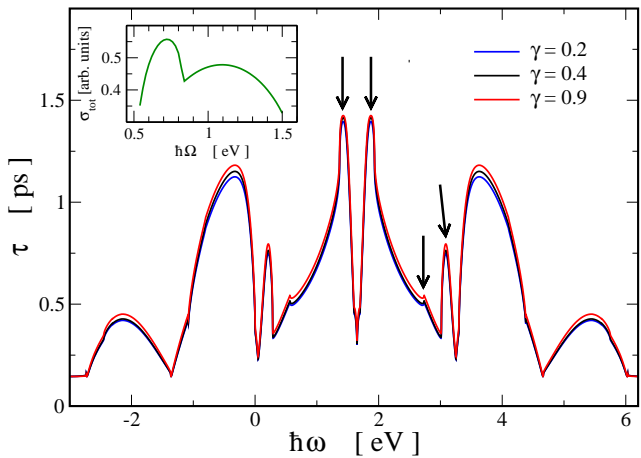}}}}
\vspace*{0.0cm}
\caption{Polaritonic lifetimes. Pump frequency $\lambda\,=\,710\,nm$, cavities
  resonance $\lambda\,=\,505.34\,nm$, refractive
  index $n = 1.97182$, pump intensity $2.4 MW/cm^{-2}$. The difference to pure
  bulk material is evident in novel, sharp features (marked) at $0..0.5\,eV$ and
  $2.5..3.3\,eV$. These energies correspond to spectrally separated
  lasing peaks in random laser systems \cite{NJP14}. Tuning the
  coupling strength of bulk matter and cavity yields small effects. {\it
    Inset:} Optical DC-conductivity of the semiconductor Mie
  resonator for $\gamma=0.2$ representing $|\chi|$$^3$ processes.}
  \label{Life}
  \end{figure}

As results for the cavities (Fig. \ref{Life})
we derive profound changes compared to pure bulk which are only in parts
expected. The nano-pillar shall support a single whispering
gallery mode. $a^{\dagger}$ ($a$) creates (annihilates) a photon as
the energy eigen-state at $\hbar\omega_0 \!\!= 2,45\,eV$
($\equiv\,505.34\,nm$). This is the {\it in-plane} Mie
resonance, which is related to the continuum of dressed electronic states
(Fig. \ref{IMAG}) and so the
non-equilibrium complex refractive index of ZnO. The momentum of the cavity photon is small compared to the
electron's momentum and $q_{\rm photon} \simeq 0$ is set whenever we consider
the electronic subsystem. We compare the lifetime for varying coupling strength
$(\!g/t)\!\!=\!\! 0.2, 0.4, 0.9$ at zero temperature and at half filling, 
yielding a suppression of the spectral function around the Fermi-level (half
width of the gap) where electrons are transferred to the high (low) energy
tails of the spectral function. Our results are derived for a pumping
intensity of $2.4 MW/cm^2$ which is typical for solid state random lasers. In agreement with results for the bulk we also find the characteristic feature
around the Fermi edge. Also the dip at $4.8\,eV$ persists in the cavity
system. The
physical interpretation of the non-equilibrium lifetimes including the Mie
resonance in (Fig. \ref{IMAG}) leads to ultrafast electro-optical Kerr effects,
non-equilibrium $|\chi|$$^3$
processes with are of the order of $0.75 ps$ up to $1.25 ps$. This is a long-living cavity polariton which couples
to the non-equilibrium AC Stark effect of the pure excited bulk. The peaks in
the lifetime of the semiconductor cavity around $2.8 eV$ and $3.2
  eV$ are occurring
comparably far from the cavities resonance frequency and so they are a proof
of the importance of the Fano resonance for electronic
spectrum in general and they express the splitting in several
  relevant polariton branches which define the lasing frequencies. This is a
  property of the independent single scatterer, which as for the random lasing
  experiments, only profits from randomness by the increased impinging
  photonic intensity. The optical DC-conductivity (inset Fig.\ref{IMAG}) of the semiconductor Mie
  resonator for $\gamma=0.2$ in the non-equilibrium, i.e. $|\chi|$$^3$
  processes in the non-equilibrium, shows a strong frequency-dependency and
  remarkable features which have not been reported before to the best of our
  knowledge. The dips may lead to significant gain narrowing.

Our results suggest, that solid state random lasers can be
  interpreted as coupled cavity polariton lasers where the Mie cavity plays a
  crucial role in the generation of topological features in the microscopic
  electronic bandstructure of the semiconductor. We theoretically demonstrated polaritonic gain in ZnO as non-equilibrium
effect in the Hubbard Model. The polaritonic states are comprised of
long-lived excitonic AC Stark states in the original ZnO gap and the Mie
resonances of ZnO nano-pillars. This implies that the ZnO laser plasma is
actually a polariton condensate which may display a threshold behavior. To distinguish between the concept
of either cavity polaritons or excitons and the particle hole plasma and
additionally the influence of the mode, i.e. the quality and the coupling of
the cavity, a temperature dependent DMFT in the the non-equilibrium will be
useful. This novel theoretical approach leads to a more fundamental
understanding of ultrafast topological effects in semiconductor
nano-cavities.\\

{\bf Acknowledgments.} RF thanks P. Guyot-Sionnest for highly efficient
discussions within the conference TIDS15 {\it Transport in
  Interacting Disordered Systems}.

\end{document}